                               \documentstyle[12pt,aaspp4]{article}
				\begin{document}     
				\title{ 
Dust in Spiral Galaxies: Comparing Emission and Absorption to Constrain Small-Scale and Very Cold Structures\altaffilmark{1}
					} 
                   
				\author{             
Donovan L. Domingue \& William C. Keel\altaffilmark{2,3}
				}
		    
				 \affil{
Department of Physics and Astronomy,     
University of Alabama,               
Box 870324, Tuscaloosa, AL 35487-0324
                                  }
\authoremail{domingue@hera.astr.ua.edu, keel@bildad.astr.ua.edu}
                                   
                                \author{
Stuart D. Ryder
                                 }
                                \affil{
Joint Astronomy Centre,
660 N. A'ohoku Place, Hilo, HI  96720
                                  }
\authoremail{sryder@jach.hawaii.edu} 

                                \author{
Raymond E. White III\altaffilmark{3}
                             }
                                \affil{
Department of Physics and Astronomy,     
University of Alabama,               
Box 870324, Tuscaloosa, AL 35487-0324
                                     }
\authoremail{white@merkin.astr.ua.edu}

\altaffiltext{1}{Based on observations with ISO, an ESA project with instruments
funded by ESA Member States (especially the PI countries: France,
Germany, the Netherlands and the United Kingdom) with the
participation of ISAS and NASA.}
\altaffiltext{2}{Visiting astronomer, Kitt Peak National Observatory, 
National Optical Astronomy Observatories, operated by AURA, Inc., under 
cooperative agreement with the National Science Foundation.}
\altaffiltext{3}{Visiting astronomer, Cerro Tololo Interamerican Observatory, 
National Optical Astronomy Observatories, operated by AURA, Inc., under 
cooperative agreement with the National Science Foundation.}

				\begin{abstract}
The detailed distribution of dust in the disks of spiral galaxies is important 
to understanding the radiative transfer within disks, and to measuring overall 
dust masses if significant quantities of dust are either very opaque or very cold. 
We address this issue by comparing measures of dust {\it absorption}, using the 
galaxy-overlap technique in the optical, with measures of the dust grains' thermal 
{\it emission} from 50-2000 $\mu$m using ISOPHOT on board ISO and SCUBA at the 
JCMT. We examine three spiral galaxies
projected partially in front of E/S0 galaxies --- AM1316-241, NGC 5545, and
NGC 5091 (for NGC 5091 we have only optical and ISO data). 
Adopting an empirical exponential model for the dust distribution, we compare 
column densities and dust masses derived from the absorption and emission techniques. 
This comparison is sensitive to the amount of dust mass in small, opaque structures, 
which would not contribute strongly to area-weighted absorption measures, and to very 
cold dust, which would contribute to optical absorption but provide only a small
 fraction of the sub-mm emission. In AM1316-241, we find global dust masses of
$2-5 \times 10^7$ $M_\odot$, both techniques agreeing at the 50\% level. 
NGC 5545 has about half this dust mass. 
The concordance of dust masses is well within the errors 
expected from our knowledge of the radial distribution of dust, and argues 
against any dominant part of the dust mass being so cold or opaque. 

The 50-2000 $\mu$m data are well fitted by modified Planck functions with an
emissivity law $\beta=-2$, at $21 \pm 2$ K; a modest contribution from 
warmer dust is required to fit only the 50 $\mu$m measurement of NGC 5545. We 
incorporate empirical corrections to the flux scale of ISOPHOT P32 data, which 
can reach a factor 2 from comparison of IRAS and ISO fluxes for objects in 
two programs.

We also present 12 $\mu$m ISOCAM observations of these pairs. The light
profiles at this wavelength exhibit shorter
disk scale lengths than in the optical. Comparison of H$\alpha$ and 12$\mu$m
images of NGC 5545 indicate that ISOCAM images are reliable 
tracers of star formation. 
				  
				\end{abstract}
						
				\clearpage
				\section{
Introduction                             
				 } 
 
Investigating the radiative ``energy-balance" within galaxies has recently become more
viable (Xu \& Helou 1996; Trewhella 1998) with the expansion of our ability 
to view their far-infrared emission. Since dust produces both 
absorption and thermal emission, its structure and distribution within spiral 
disks affect how that radiation is released. 
Both the emission properties of interstellar dust grains and the total mass in 
grains determine how stellar (optical-UV) light is reradiated as the total infrared flux.
If dust exists in very clumpy structures (behaving as large grains with
inefficient absorption), inferring dust masses from area-weighted extinction 
measures would yield
underestimates. If a large amount of cold dust exists with FIR emission below
the detection limits of ISOPHOT (Lemke et al. 1996), then modified 
blackbody temperature fitting will yield mass underestimates, as IRAS was shown
to have done by using ISOPHOT data evaluate dust masses (Alton, et al. 1998
; Haas et al. 1998; Kr\"{u}gel et al. 1998). 
The latter claim is that using IRAS data alone leads to
underestimates of dust masses by a factor of 10, because of the IRAS 100 $\mu$m
cutoff and the contribution of warm dust to the 60 $\mu$m measurement. Block
at al. (1994) demonstrated that radiative transfer models with $B$ and 
$K^\prime$
imaging can increase the dust mass estimates of two spiral disks by an order of
magnitude over IRAS values.

Debates over the opacity of spirals extend from statistical studies by 
Holmberg (1958) and Valentijn (1990), using the inclination-surface-brightness
tests, to reassesments of that work by Burstein, Willick, \& Courteau (1995) 
and Huizinga (1994).
Claims have ranged disks being opaque to being transparent.
``Energy-balance" methods (Evans 1995; Xu \& Helou 1996) have relied solely on 
IRAS data with a limiting upper wavelength of 100 $\mu$m, while the peak of FIR
emission is expected to be between 100 and 300 $\mu$m (Evans 1992). 
Campaigns undertaken to directly measure the opacity within spiral galaxies
(White \& Keel 1992; Keel \& White 1995; White, Keel \& Conselice 1996, 1999;
;Gonzales et al. 1998; Domingue, Keel, \& White 1999)
have shown that moderate extinctions predominate
in disks, with spiral arms exhibiting higher levels than interarm regions. 
These methods exploit the unique opportunity afforded by partially overlapping 
galaxies. By using the measured extinctions in the $B$ band and converting them
to column densities we deduce the total mass of light absorbing dust grains in
such overlapping pairs. 

A comparison of dust emission and absorption measures is
informative since the two complementary techniques have
very different sensitivity to dust temperature and clumping. Absorption
measures are insensitive to grain temperature, but the mean absorption
will be reduced for a given dust mass if the dust is strongly clumped. Conversely,
emission measurements are insensitive to dust distribution (as long as
the mid- and far-IR optical depth remains small), but they are sensitive
to dust temperature, varying as $\sim$$T^{-5}$ for a fixed wavelength of
observation. The derived masses in both cases scale as ${{\rm H}_0}^{-2}$, so 
their ratio is distance-independent.

We observed three overlapping galaxy pairs using ISOPHOT mapping from 
50-200 $\mu$m, along
with SCUBA observations up to 2000 $\mu$m for two members of the sample. 
The far-IR measurements were supplemented with 12$\mu$ ISOCAM mapping,
to give some insight into the structure of at least the warmer dust, and
seek evidence of obscured star formation in the strong dust lanes in these
spirals. These
data allow us to fit temperatures to the FIR emission and calculate total dust 
masses
based on FIR emission models (Hildebrand 1983; Rowan-Robinson 1992). We then compare
dust masses from these two independent methods (FIR/sub-mm emission and directly 
measured extinction)  to determine the role of dust's large scale structure
in spiral disks.
 
				\section{
Observations and Data Reduction  }

			       \subsection{
Sample Selection                 }

The three overlapping-galaxy pairs for this study were chosen from the
large survey performed by White, Keel, and Conselice (1999), which surveyed
candidates from numerous catalogs to find pairs with the appropriate geometry
and symmetry for analysis of the absorption. We believe them to be the
apparently largest and brightest partially overlapping spirals within the
area of sky accessible to the ISO primary mission.

The proximity in both position and velocity for these pair members virtually
guarantees that they are gravitationally bound, either as pairs or as members
of small groups. This might introduce the possibility that the spirals
we observed are atypical because of interactions. In selecting the sample
for this study, we attempted to reduce such effects by eliminating objects
showing morphological evidence for tidal distortion, and excluding cases
where the presence of a starburst gave hints of externally triggered star
formation. Specifically, aside from lack of tidal features in our optical
imagery, we can note that the global H$\alpha$+[N II] equivalent width in 
NGC 5545 (30 \AA\ ) is very close to the median level found for noninteracting 
Sc galaxies (Kennicutt et al. 1987). Neither our ground-based nor {\it HST}
imagery shows unusual levels of disk star formation, as traced by bright 
blue knots. Earlier reports of NGC 5091 as ``tidally disrupted" (e.g., Smith
\& Bicknell 1986) appear to rely on interpretation of the NW arm, from
photographic plates, as detached from the disk and thus tidally induced. CCD 
imagery shows that the apparent gap results from the prominent and extensive
dust lane crossing the disk in this region, so that this arm structure is
not in itself evidence for strong tidal interaction. The radio fluxes of
the spirals are all in typical ranges for the type, when detected at all;
there is a weak detection of the nucleus of NGC 5545 (Hummel et al. 1987),
while AM1316-241 is undetected in the PMN survey (Griffith et al. 1994) and
NGC 5091 does not appear in radio maps of NGC 5090 (which is a classical
twin-jet radio galaxy, as noted by Lloyd, Jones, \& Haynes 1996).
Finally, all these spirals have rather cold colors across the
50-200$\mu$ range, another indication that any triggered burst of star
formation must be weak in comparison with the steady-state rate. Our intent
was to select the most typical spiral galaxies for which this kind of analysis
can now be performed. Given the connection between pair membership and the
incidence of grand-design spiral patterns (Elmegreen \& Elmegreen 1982), 
and the global spiral patterns
shown by each of these galaxies, our results can strictly apply only to 
grand-design spirals, and it is in this sense that we feel our results may
be significantly affected by the proximity of the background galaxies. 
Such an emphasis is almost unavoidable in using the overlap technique to measure
dust extinction.

			       \subsection{
Mid-Infrared Observations
				}
Three pairs of galaxies (see Table 1) were observed with ISOCAM (Cesarsky et al.
1996) on board the
$Infrared$ $Space$ $Observatory$ (ISO) (Kessler et al. 1996) with the LW10
filter(8-15 $\mu$m), which
was designed to match the IRAS 12 $\mu$m filter. Each pair was observed
in a 2 $\times$ 2 raster mode with a 3$\arcsec$ pixel scale. Steps of
6$\arcsec$  
were used between each pointing with integration times of 10 seconds each. The final images 
were 34 $\times$ 34 pixels.
The data were reduced with CAM Interactive Analysis (CIA) for dark subtraction,
transient removal, flat-fielding and co-addition of frames. The small rastering
step size relative to the pairs' apparent sizes did not allow for the production
of flat fields from the data. It was necessary to use the flats from the
calibration library which can vary from true flats due to changes in the 
position of the filter wheel (Siebenmorgen et al. 1996). Median background 
levels were determined from histograms of the outlying regions (see Table 2). 
The resulting images are in figures 1 through 3. 

All six galaxies are detected in the 12 $\mu$m LW10 filter. The elliptical in
AM1316-241 is unresolved with a peak to total flux ratio of 0.21, comparable 
to the 
LW10 PSF with 3$\arcsec$ pixels. NGC 5544 is nearly unresolved
and all other galaxies in this sample are clearly not point sources. NGC 5091 did not entirely fit into the chosen CAM field-of-view. Its flux
was modeled by compositing the image with the symmetric counterpart mirrored
about the galaxy's center. 

To obtain information on the warm dust distribution, we did conventional ellipse
fitting and produced exponential disk models to convolve to the ISOCAM LW10 PSF.
Best (two-dimensional) fits to the ISOCAM data for the three spirals are 
used to 
estimate 12 $\mu$m exponential scale lengths (see table 2). 

			      \subsection{
Far-Infrared Observations
			      }
The three pairs were observed using the ISOPHOT instrument onboard ISO in P32
mapping mode at 50,100, and 200 $\mu$m (see Figure 4). A 3 x 3 rastering with an
oversample factor of 3 yielded maps with pixels of 15$\arcsec$ x 15$\arcsec$ . The
data were reduced using PHT Interactive Analysis (PIA). Each observation
included two
Fine Calibration Source (FCS) measurements, one before and after pointings
at the target. This was to correct for any detector responsivity fluctuations 
within the time frame of each observation. Histograms of the resulting images
were used to obtain the background level used for subtraction. 

Neither AM1316-241 nor NGC 5090/1 was detected at 50 $\mu$m. A possible
explanation for non-detection is the C100 detector's transient behavior as
explained below. These two pairs are detected at 100 and 200 $\mu$m and NGC 5545
is detectable at all three wavelengths. 

P32 is not currently one of the validated modes of ISOPHOT. This is
due to the present inability to calibrate its chopped rastering (Klaas et al.
1998). We apply corrections to calibrate fluxes
from this set of data through
statistical comparison with IRAS fluxes of these pairs. The present difficulty
involves the transient behavior in the C100 detector. Fluxes are underestimated
relative to the background because the time scale of the C100 detector to react 
to the illumination step is comparable to the time between chopper pointings
(Tuffs et al. 1996; Laureijs 1999). For point sources this will have a 
significant affect on the total flux. Our background surface brightness 
measurements are found to be consistent with the 30\% overestimation figure
quoted by Alton et al.(1998) in relation to COBE estimations from IRSKY of the 
same sky backgrounds. This does not mean our fluxes are overestimated, which
may differ from the situation for fully resolved targets (Alton et al. 1998).
To assist in our calibration efforts we used 20 additional ISOPHOT C100 
images at 60 and 100 $\mu$m of pairs from the Karachentsev Catalog (KPG), taken in an 
observing program by J. W. Sulentic. All of these fluxes
were contrasted to IRAS ADDSCAN fluxes at 
60 and 100 $\mu$m for the KPG pairs (Toledo 1998) and our own sample 
(see Figure 5). A linear fit to the results shows ISOPHOT C100 fluxes to be a
factor of 2.2 higher than IRAS fluxes with a small additive offset, which
differs for the groupings below and above a measured ISOPHOT flux of 1 Jansky.
We apply the multiplicative factor correction of 2.2 to our ISOPHOT
C100 fluxes (see Table 3). The affect of transient behavior on the C200
detector is assumed to be smaller (Laureijs 1999). To find a reasonable
correction for the C200 data we incorporate our SCUBA flux estimates of
AM1316-241 and NGC 5545. Modified blackbody fits to our corrected C100 fluxes and SCUBA 
flux limits indicate that our C200 data do not require any corrections
to be consistent with submillimeter and millimeter observations. 

			     \subsection{
Submillimeter and Millimeter Observations
			     }

We observed AM1316-241 and NGC 5544/5 from 450 $\mu$m to 2 mm with the
SCUBA instrument (Holland et al. 1999) at the 15m James Clerk Maxwell 
Telescope (JCMT) on Mauna Kea. The measurements were made on 11 February 1999, with a 
typical atmospheric $\tau=1.5$ at 450 $\mu$m. The two independent bolometer arrays were used for 
simultaneous 450 and 850 $\mu$m mapping, in the 64-point mode yielding full
spatial coverage at both wavelengths. The single-pixel 1.3- and 2-mm
detectors were used in a chopped photometric mode centered on each spiral 
component, with sky chopping at a throw of 120\arcsec set to miss the optical
and far-infrared extent of the pair members. Flux calibration used Mars, with
the 
same observing modes as for the galaxy data. The mapping fields were centered 
between the members of each pair, by offset after checking the pointing on AM1316-241 using 
3C 279, which also served as a point reference source for mapping the 
telescope beam. The pointing check on NGC 5544/5 was 1308+326.

In constructing the $450/850 \mu$m maps, noisy bolometers and corrupted
integrations were flagged and subsequently ignored. The unflagged measurements 
were gridded and interpolated into 1$\arcsec$ pixels in the $\alpha,\delta$ (tangent) 
plane, using the rotation of the altazimuth telescope coordinates on the sky 
to fill in blank areas from noisy detectors. Using these maps at 10$\arcsec$
resolution 
gave a $3-\sigma$ detection of the spiral in AM1316-241 at $21 \pm 7$ mJy at 
850 $\mu$m (at the expected coordinates, giving confidence in the detection), and 
a 450 $\mu$m limit of 690 mJy. For NGC 5545, we obtain $3 \sigma$ upper limits 
of $< 31$ mJy (850 $\mu$m) and $<450$ mJy (450 $\mu$m).
   
The spiral members of these pairs are not well resolved by the JCMT beam at
millimeter wavelengths, taking either the optical or ISO imagery as a guide 
to the extent of the emitting dust. The FWHM of the beam is about 31$\arcsec$
and 48$\arcsec$ 
in the two bands, while the scale lengths for exponential-disk fits to the 
ISOCAM $12 \mu$ images are 2.5$\arcsec$ for AM1316-241 and 8.0$\arcsec$ for NGC 5545. Accordingly,
the measurements at 1.3 and 2 mm used the SCUBA ``photometry" detectors in 
point-source fashion. The following 3$\sigma$ upper limits were derived: for 
AM1316-241, F(1.35 mm) $<$7.8 mJy, F(2 mm) $<$ 27 mJy; for NGC 5545, 
F(1.35 mm) $<$12.8 mJy and F(2 mm) $<$ 35 mJy. ISO mapping has shown that for 
many spirals, scale lengths measured in the optical and mid-IR are useful 
proxies (at the $\sim 30$\% level) for the far-IR distribution (Alton et al. 
1998, Haas et al. 1998), so we calculated 1.3-2.0 mm beam corrections based 
on the structures both at optical wavelengths and at $12 \mu$, multiplying 
these detection limits by the larger of the optical or mid-IR corrections 
before using them in analysis. These beam corrections are scaling factors
of 1.30 (1.15) at 1.35 (2.0) mm for AM1316-241 and 1.5 (1.25) for NGC 5545.
The resulting global limits are stringent enough to let us rule out 
significant emission from very cold dust. 

                               \subsection{
Supporting Optical and Near-Infrared Imaging
                                }

We adopt the absorption measures for AM1316-24 presented by White
\& Keel (1992). For NGC 5545, we have done a similar analysis using
$B$ and $I$ imagery obtained with the prime-focus CCD imager at the
4m Mayall telescope of Kitt Peak National Observatory. The background
system NGC 5544 was modelled by fitting elliptical isophotes to the
non-overlapping regions and interpolating across the overlap area.
The galaxy is slightly asymmetric, so that the usual reflection/subtraction
technique employed for elliptical galaxies
gives much larger (and systematic) residuals. 
To subtract light from the foreground system, we employed rotational
symmetry after applying a 6$\arcsec$ median filter to remove bright associations
so that residuals due to individual bright clusters and stellar associations 
would not produce spurious absorption regions.
The resulting maps of residual intensity are shown in Fig 6. For
regions not disturbed by bright foreground structures, we derive
mean values for the interarm region at r=29$\arcsec$ $A_B=0.52$,$A_I=0.30$,
and for the prominent arm crossing this radius, $A_B=1.3, A_I=1.0$.
The radial profile of $B$-band light implies a disk scale length of
12$\arcsec$, so these measurements apply 2.4 scale lengths from the nucleus.

For NGC 5090/1, we use a $B$-band image obtained with the CTIO 
1.5m, to estimate extinctions within NGC 5091, 
starting with our usual practice of modelling and subtracting the
light distribution of the background elliptical. The prominent dust
lane on the NE side of NGC 5091 in fact goes very close to zero intensity
after this exercise, demonstrating that the spiral is indeed in front.
The typical second step - modelling and subtracting the spiral's own
light by symmetry - does not work well in this case, producing unphysical
negative flux in much of the overlapped disk. The strong dust lane has
no symmetric counterpart, so we are driven to a more uncertain estimate based 
on the minimum residual intensity in other parts of the disk lane,
and just inside its inner edge, at other points with similar radius in the
disk plane but essentially no background light. This yields a lower
limit $A_B> 2.8$ at the deepest part of the dust lane, about $A_B=0.6$
at a radius of 19", and $A_B < 0.2$ outside the dust lane at $r=38\arcsec$.
The disk has substantial local structure, but a compromise exponential
scale length of $13.5\arcsec$ should be adequate for a dust-mass estimate.

Global estimates of dust mass for these pairs assumed that equal amounts
of dust are in the arm and interarm regions, as suggested by the surface 
areas and extinctions in the areas that we have measured.

				 \section{
Results                          }

                                \subsection{
Mid-Infrared 
                                }

The ISOCAM 12 $\mu$m images show us very warm dust emission in the three pairs.
The spirals in AM1316-241 and NGC 5544/5 dominate because they are expected to 
contain more dust and have more gas available for star formation. The elliptical
dominates over the spiral in NGC 5090/1, but this early-type galaxy is also a 
radio galaxy, which suggests an alternate heating source for the dust other than
the interstellar radiation field. A convolution
of an H$\alpha$ image of NGC 5545 (Kennicutt et al. 1987) to the ISO LW10
resolution reveals a very good correlation of emission features at these two wavelengths (see Figure 7).
This suggests that virtually everything we see in the ISOCAM map is reliable and ties the mid-IR
structure to star formation, since H$\alpha$ is a good indication of 
star-formation rate (SFR) (Kennicutt 1983). ISOCAM observations of other spirals (Wozniak et al. 1998; Charmandaris et al. 1999) show correlations in the
comparison of H$\alpha$ and mid-IR images. 
In AM1316-241 and NGC 5544/5 the mid-IR emission scale lengths are shorter than 
those of the $B$-band indicating a central predominance of star formation, 
although the outer disk in NGC 5091 does contain mid-IR emission (Figure 2). 
			        \subsection{
Extinction Dust Masses
			     }
       
Using radial extinction data for spiral disks, White, Keel \&
Conselice (1999) empirically show that a declining exponential (with the same scale
length as the $B$-band disk starlight) is a good model
for dust distribution in the interarm regions of spirals with 
about half of a galaxy's
dust being located within spiral arms. We calculate column densities of
localized regions with measured extinctions and integrate these over the 
entire disk, adopting the empirical exponential form to derive the total dust
mass as seen in extinction. 

To generate column densities from the extinction values, we must know the 
absorption cross section
of dust grains at the relevant wavelengths. The size distribution of grains
is weighted toward the larger particles of 0.1-0.2 $\mu$m (O'Donnel \& Mathis
1997) with a classical mean size chosen 
as 0.1 $\mu$m (Hildebrand 1983) and overall density
of 2.8 g cm$^{-3}$. This density is based on an average density of the actual 
predicted constituent particles of the ISM -- graphite having a density of 2.26 
g cm$^{-3}$ and silicates such as olivine having densities of 3.3 g cm$^{-3}$
(Draine \& Lee 1984). 
Another model (O'Donnel \& Mathis 1997) argues that the larger grains
($> 0.02 \mu$m) are composite porous structures with a lower density of
$\sim$ 1.3 g cm$^{-3}$. Mass estimates are therefore sensitive to the possible 
fractal nature of grains. Since 0.1 $\mu$m grains are part of this possibly
porous group, we use the porous grain value of 1.3 g cm$^{-3}$, while noting that masses may be underestimated by a factor of $\sim$2 . For particles of this size, absorption 
efficiencies were calculated by Draine \& Lee (1984) using synthesized
dielectric functions of ``astronomical silicate" and graphite. The efficiencies
$Q_{abs} \equiv C_{abs}$/$\pi$a$^{2}$ yield graphite cross sections $C_{abs}$ of 1.6$\times10^{-11}$ cm$^{2}$ and 6.0$\times10^{-11}$ cm$^{2}$ for the $I$ and
$B$-bands respectively. Silicate values are 1.3$\times10^{-10} cm^{2}$ and 
2.3$\times10^{-10} cm^{2}$ for the $I$ and $B$-bands. Silicate and graphite models 
,in conjunction
with observational evidence, suggest that these species are mixed in
a ratio of 1:1 (Mathis, Rumpl, \& Nordsieck 1977; Draine \& Lee 1984). 
The cross sections we adopt are the average of the graphite and silicate
values. The localized apparent column density is
\begin{equation}
N(r)= {\tau(r)\rho \over C_{abs}},
\end{equation}
where $\tau$ is our value for the opacity in that region and $\rho$ is the
density of the grains, chosen to be 1.3 g cm$^{-3}$. The central value is
\begin{equation}
N_{0} = N(r)e^{r/r_{0}},
\end{equation}
derived from our exponential model where $r_{0}$ is the scale length of the disk's 
$B$ image. We assume the long wavelength emission to have the same distribution
as that of the $B$-band (consistent with the results of Alton et al. 1998). 
 
The total dust mass is found through the integral
\begin{equation}
M_{d} = 2\pi(b/a)N_{0}\int_{0}^{\infty}e^{-r/r_{0}}rdr,
\end{equation}
which becomes
\begin{equation}
M_{d} = 2\pi(b/a)N_{0}r_{0}^{2},
\end{equation}
where $b/a$ is the axial ratio of the galaxy. We have taken $H_{0}=75$ 
km s$^{-1}$ Mpc$^{-1}$. Interarm optical depths $\tau_{B}$ and
$\tau_{I}$ of AM1316-241 are 0.3 and 0.2 (both at 2.4 $r_{0}$) respectively, while for NGC 5545,
$\tau_{B}$ is 0.48 (at 2.4 $r_{0}$), and $\tau_{B}$ is 0.55 (at 1.4 $r_{0}$) for NGC 5091. A final multiplication of the mass by a factor of two is
required because we assume that interarm dust 
represents only half of the dust mass,
with the other half in the spiral arms (which fits with the direct
extinction measurements cited in the Introduction).
Dust masses (see Table 4) are derived from these quantities are 
$1.6\times10^{7} M_{\odot}$ in NGC 5545, 3.9$\times10^{7} M_{\odot}$ in 
AM1316-241 and 0.8$\times10^{7} M_{\odot}$ in NGC 5091 using the $B$-band opacities. 
The $I$-band estimateof the dust mass in AM1316-241 is 4.5$\times10^{7} M_{\odot}$, consistent with the $B$-band determination.  
				 \subsection{
FIR Dust Masses                           }

We consider the most important component of the dust emission by mass, which
is the coolest component that is well represented in the spectrum. To a
good approximation, we find that the $100-2000 \mu$ data can be fitted to
a single modified blackbody law in each case, with a hotter component
(associated with star-forming regions) required only to account for
the 50 $\mu$m emission in NGC 5545.

Single-temperature fitting to the FIR fluxes, with an 
emissivity law of $\nu^{\beta} B_{\nu}(T_{d})$ with $\beta=2$, gives us the total
dust mass within these spirals provided there is no significant colder
component of their ISM. This modified blackbody emissivity function is a good
representative of the FIR emission of dust grains. We choose $\beta=2$ because
it has been shown to fit the spectrum of ISO data on spirals (Kr\"{u}gel et al. 1998) and our limited number of data points are not suited to constraining 
this parameter. Temperature errors were estimated using a Monte Carlo
approach, the modified
blackbody fits were applied to points with gaussian distributions about the 
measured fluxes using an error of 30\% for ISOPHOT data (as in Table 3) and the SCUBA detection error at 850$\mu$m in the case of AM1316-241 (Table 3). This gives us a mean temperature and 1-$\sigma$ errors (see Table
4).
In AM1316-241 we are able to fit fluxes at 100, 200, and 850 $\mu$m.
NGC 5091 and NGC 5545
have fits only to 100 and 200 $\mu$m data points with very useful upper limits 
at sub-mm and millimeter wavelengths for NGC 5545. The FIR flux in these three
pairs is dominated by the foreground spiral since the background galaxy in each
case is an early type E or S0 (see table 1) with the possible exception of 
NGC 5090/1 in which the elliptical is a radio galaxy that is dominant at $12\mu$m. The dominance of spirals in the IR
is expected as a result of early-type galaxies' deficiencies in gas and dust
content and observationally confirmed for IR luminous pairs of different Hubble
types (Bushouse, Telesco, \& Werner 1998) and our own ISOCAM observations,
excluding NGC 5090/1, presented in this paper.  
Resulting temperatures of 21 K therefore represent those of a cold component
of the dusty
spirals in AM1316 and NGC 5545. Temperatures derived from IRAS 60 $\mu$m and 100
$\mu$m fluxes are 29 K and 25 K respectively. The derived temperature for
NGC 5091 is 22 K, which agrees with the IRAS temperature within the errors.

An additional warmer component of dust is required to fit the 50 $\mu$m flux of
NGC 5545. We add a 47 K dust population which keeps the added flux within
the errors and fits both ISOPHOT $50\mu$m and IRAS 25 $\mu$m data points
(See Figure 8). The upper limit on the AM1316 $50\mu$m emission allows for 
a similar warm dust component, but the possible contribution of the early-type
member of the pair to the IRAS 25 $\mu$m flux would cause the derived temperature
to be too high. If none of the emission is from
the elliptical, we can impose a strict upper limit of 60 K to this warm
dust. The masses of the 40 K and above components as calculated in the manner
below are only 1\% of the total dust mass and are relatively insignificant. 
The temperature fit in NGC 5545 strongly depends on the reliability of the
200 $\mu$m flux. A detection at 850 $\mu$m would better constrain the fit below
the upper limit at that wavelength. 

Interpreting the FIR emission as a total dust mass requires a understanding of
the grain composition within spiral disks. We consider two models for this 
emission.
The simple single grain model for
mass determination (Hildebrand 1983) is:
\begin{equation}
M_{d} = {4a\rho D^{2} \over 3} {F_{\nu} \over Q_{\nu}B_{\nu}(T_{d})},
\end{equation}
where $\rho$ = 3 g cm$^{-3}$ is the grain density, $a = 0.1 \mu$m is the grain
size, D is the galaxy distance ($H_{0}=75$ km s$^{-1}$ Mpc$^{-1}$)
, $F_{\nu}$ is the measured flux, $B_{\nu}(T_{d})$ is the Planck function, and
$Q_{\nu}$ is the grain emissivity where
\begin{equation}
Q_{\nu} = {3 \over 1300}({\nu \over \nu_{125\mu m}})^{2},
\end{equation}
and $\nu_{125\mu m}$ is the frequency corresponding to $125\mu$m.
This model has been applied to the 200 $\mu$m ISOPHOT fluxes and 100 $\mu$m
IRAS fluxes of several nearby spiral galaxies by Alton et al. (1998) to estimate
masses at 3 - 10$\times10^{7} M_{\odot}$. When applied to 
the three pairs our sample, this model predicts dust masses (see table 4) of 
0.1 - 2.6$\times10^{7} M_{\odot}$. 

A more complex model (Rowan-Robinson 1992), which takes into account different 
grain sizes and distributions, is based on the behavior of the infrared 
cirrus. This cirrus model separates cirrus and starburst components in galactic
emission and can be well fit by the $\nu^{2}B_{\nu}(T_{d})$ emissivity. The 
parameters of this model require a temperature and a 100 or 800 $\mu$m flux
along with the distance of each galaxy. The scaling factor
log(M$_{d}$/$S_{100}D^{2}$) is approximately 3 for temperatures of 21
- 22 K. We derive dust masses (see Table 4) which are similar to those of the
Hildebrand (1983) model and range from 0.1 - 3.2$\times10^{7} M_{\odot}$.

                                   \section{
Summary and Discussion
				}

Far-infrared photometry from 50-2000 $\mu$m, when used together with extinction
maps of spiral disks, reveals the nature of the dusty structures within those
galaxies.
Grain models, plus ISOPHOT FIR and SCUBA 
sub-mm/mm observations yield dust masses of 1-3$\times10^{7} M_{\odot}$ for NGC 5545 and 
AM1316-241.
Extinction measures in these overlapping spiral galaxies lead to comparable mass estimates.  

In NGC 5545, our sub-mm flux limits restrict any significant
mass of undetected dust to $T<$7 K, while $T<$10 K for any undetected dust in
AM1316, i.e. the dust mass of
these two galaxies can only increase by a factor of 2 or more if there exists a
significant contribution below these temperatures. A lack of evidence from COBE
for grain temperatures below 13 K in the Milky Way (Lagache et al. 1998) argues 
for an upper limit of twice the tabulated dust mass unless these galaxies
contain a very different type of dust grain population than the Milky Way. 

In NGC 5091, the large difference in dust mass resulting from the two methods 
permits a larger fraction of cold dust to exist, but the incomplete data set 
for this pair
limits the interpretation of the results and the possible far-IR contamination 
by the radio elliptical may affect the temperature fitting. 

Because the dust masses derived from extinction measures are slightly higher
than those
derived from FIR emission, dust cannot be hiding in clumpy
structures. Small opaque structures would not have a strong affect on
area weighted absorption measures. In such a case, absorption measures would be interpreted as lower dust masses than FIR models would predict. We do not
see any evidence for this possibility and it is unlikely that any significant
amount of dust exists in opaque structures. Small, opaque dust structures at $T<10$ K
cannot be ruled out, but there seems no physical motivation for that option.

Given the uncertainties in each of these calculations, their agreement is
striking, and
suggests that both techniques observe essentially all the dust in typical spirals, in as far as these gravitationally pair-bound spirals can be considered
typical of grand-design spiral galaxies.

We thank G. Moriarty-Schieven for assistance with the JCMT observations. 
The ISOPHOT data presented in this paper were reduced using ``PIA", which is a joint development by the ESA Astrophysics Division and the ISOPHOT consortium.
The ISOCAM data presented in this paper were analysed using ``CIA", a joint 
development by the ESA Astrophysics Division and the ISOCAM Consortium. The
ISOCAM Consortium is led by the ISOCAM PI, C. Cesarsky, Direction des Sciences de la Matiere, C.E.A., France. D.D. greatly appreciates the training received
using these data analysis packages while attending the Les Houches summer 
school entitled "Infrared Space Astronomy Today and Tomorrow". We thank J.W.
Sulentic for the use of additional ISOPHOT data to assist in P32 corrections.
This work was supported by NASA/ISO grant NAG 5-3336.

				  
				\clearpage
				\title{       
Figure Captions                               
				}


                               \figcaption{
$B$ band HST image of AM1316-241 (left) and ISOCAM LW10 image (right) of the
same field. The elliptical is a point source to ISOCAM.
				}


                               \figcaption{
$B$ band CTIO image of NGC 5090/1 (left) and ISOCAM LW10 image (right) of the
inset field-of-view. The foreground disk exhibits low emission.
				}

                               \figcaption{
$B$ image of NGC 5544/5 (left) with masked background and ISOCAM LW10 image (right) of the same field. The SB0 (NGC 5544) only appears as a faint nucleus 
at 12 $\mu$m. The interarm separation in the North-East spiral arm is noticable in the ISOCAM image. 
		}

                               \figcaption{
ISOPHOT P32 maps of (left to right) AM1316, NGC 5090/1, NGC5544/5 at (top to bottom)
50, 100 \& 200 $\mu$m. 
				}


                               \figcaption{
A comparison of IRAS flux estimates vs. fluxes from 20 ISOPHOT P32 maps of 
pairs at 60 \& 100 $\mu$m. Plus signs \& asterisks represent pairs from an
ISOPHOT observing program by J.W. Sulentic
at 60 \& 100 $\mu$m respectively. Diamonds represent the three pairs in this
paper which were observed at 100 $\mu$m. The best fit slope is 2.2, suggesting
the needed correction to these ISOPHOT data.
				}	


                               \figcaption{
Residual intensity maps (e$^{-\tau}$-1) in $B$ (top) and $I$ (bottom) of the overlap region in NGC 5544/5. Dark regions indicate higher extinction. 
 }


                               \figcaption{
An H$\alpha$ image of NGC 5545, the same image convolved to the PSF of ISOCAM LW10, and the ISOCAM LW10 image for comparison. Every structure in the CAM
image can also be seen in the convolved image including the N-E arm, S-W arm 
and the SB0 nucleus.
                 }


                               \figcaption{
The temperature fits of AM1316-241 and NGC 5545. ISOPHOT (50, 100, 200 $\mu$m) 
data and SCUBA JCMT (450-2000 $\mu$m) data are represented by boxes and IRAS 25 $\mu$m data are diamonds. The sum of the 47 K and 21 K fits to the spectrum of
NGC 5545 is displayed as the dashed line.
                               }

				\end{document}